\documentclass[10pt,letterpaper,conference]{IEEEtran}

\usepackage{cite}
\usepackage{amsmath,amssymb,amsfonts}
\usepackage{algorithmic}
\usepackage{graphicx}
\usepackage[dvipsnames]{xcolor}
\usepackage[final]{microtype}
\usepackage[T1]{fontenc}
\usepackage{textcomp}
\usepackage[all]{nowidow}
\usepackage[keeplastbox]{flushend}

\usepackage{booktabs}
\usepackage{multirow}
\usepackage{xspace}
\usepackage{url}
\usepackage{balance}
\usepackage[colorlinks=true,linkcolor=magenta,citecolor=cyan]{hyperref}
\usepackage{subcaption}
\usepackage{makecell}     
\usepackage{comment}      
\usepackage{tikz}
\usetikzlibrary{arrows.meta, positioning, fit, calc, shapes.geometric,
                shapes.symbols, backgrounds, decorations.pathreplacing,
                decorations.markings}
\usepackage{enumitem}     
\usepackage{listings}     

\lstdefinestyle{cmdbox}{
  basicstyle=\ttfamily\footnotesize,
  backgroundcolor=\color{black!6},
  frame=single,
  framerule=0.3pt,
  rulecolor=\color{black!35},
  xleftmargin=4pt, xrightmargin=4pt,
  framexleftmargin=4pt, framexrightmargin=4pt,
  aboveskip=5pt, belowskip=10pt,
  breaklines=true,
  postbreak=\mbox{\space\space},
  columns=fullflexible,
  keepspaces=true,
  language={},
}
\lstnewenvironment{cmd}{\lstset{style=cmdbox}}{}

\newcommand{\todo}[1]{}

\title{Hydra: Phase-Aware Workload Characterization of LLM Inference across Edge SoC Generations, Backends, and Quantization Levels}

\author{%
\IEEEauthorblockN{Amir Taherin\IEEEauthorrefmark{1},
Sana Taghipour Anvari\IEEEauthorrefmark{1},
Charles Amante\IEEEauthorrefmark{1},
Yixiao Chen\IEEEauthorrefmark{1},
Ruben Noroian\IEEEauthorrefmark{1},
Zlatan Feric\IEEEauthorrefmark{1},\\
Nicolas Bohm Agostini\IEEEauthorrefmark{1},
Pu Zhao\IEEEauthorrefmark{1},
Jos\'e Cano\IEEEauthorrefmark{2},
Bin Ren\IEEEauthorrefmark{3},
Yanzhi Wang\IEEEauthorrefmark{1},
David Kaeli\IEEEauthorrefmark{1}}
\IEEEauthorblockA{\IEEEauthorrefmark{1}Northeastern University, Boston, MA, USA \quad
\IEEEauthorrefmark{2}University of Glasgow, UK \quad
\IEEEauthorrefmark{3}College of William \& Mary, VA, USA\\
\{taherin.a, taghipouranvari.s, amante.c, chen.yixia, noroian.r, feric.z,\\
bohmagostini.n, p.zhao, yanzhiwang, d.kaeli\}@northeastern.edu\\
Jose.CanoReyes@glasgow.ac.uk, bren@cs.wm.edu}
}

\begin{document}
\maketitle
\begingroup\renewcommand\thefootnote{}\footnotetext{\copyright~2026 IEEE.
Personal use of this material is permitted. Permission from IEEE must be
obtained for all other uses, in any current or future media, including
reprinting/republishing this material for advertising or promotional
purposes, creating new collective works, for resale or redistribution to
servers or lists, or reuse of any copyrighted component of this work in
other works. Accepted at IEEE IISWC 2026.}\endgroup

\begin{abstract}
Edge LLM deployment is shaped by more than model size and precision: inference backend, hardware platform, memory traffic, and power management all affect latency and efficiency. We present \emph{Hydra}, a common-schema, phase-aware workload characterization framework for LLM inference on edge SoCs. Hydra instruments HuggingFace Transformers and \texttt{llama.cpp} with a shared per-prompt timing schema and fuses those records with hardware telemetry,
enabling a multi-dimensional characterization of performance, system-resource utilization, and efficiency across prefill and decode phases. Using Hydra, we evaluate three consecutive edge System-on-Chip (SoC) generations
(AGX Xavier, AGX Orin, and AGX Thor), 13 instruction-tuned LLMs from seven
families, five execution formats, and consider input/output-length sensitivity. The resulting artifact contains roughly $107$K per-prompt records and is publicly released with Hydra. Our analysis shows that aggregate latency alone hides key deployment effects: backend structure changes where latency is introduced, quantization reduces memory traffic and energy but does not predict power monotonically, and SoC generation changes how utilization and efficiency should be interpreted. By connecting phase-level timing with system-resource utilization and efficiency metrics, Hydra enables reproducible, phase-aware characterization of edge LLM inference. Hydra's source code and the collected per-prompt trace corpus are available open-source at: \url{https://github.com/amirtaherin/hydra}.
\end{abstract}


\section{Introduction}
\label{sec:introduction}

Large language models (LLMs) have transitioned from research artifacts to production-critical infrastructure, powering code generation~\cite{chen2021evaluating}, machine translation~\cite{costa2022no}, conversational agents~\cite{zhang2019dialogpt,openai2023chatgpt,anthropic2023claude}, and a growing set of multi-modal applications~\cite{rupprecht2026survey} served at scale by modern inference stacks~\cite{kwon2023vllm,patel2024splitwise,zhong2024distserve,xiang2025servegen}. 
The dominant deployment standard remains datacenter serving: model weights, a key--value (KV) cache, and per-token computation reside on discrete server-class accelerators. These systems utilize high-bandwidth memories (HBM) and abundant power and thermal headroom. Yet a growing class of latency-sensitive and privacy-sensitive applications such as robotics~\cite{zeng2023large,taherin2026glsvlsi,lin2026vote}, autonomous driving~\cite{cui2024survey,wang2024survey}, healthcare monitoring~\cite{mohammed2023smart}, smart homes and on-device agents~\cite{vardakis2024review} cannot always tolerate the round-trip latency, privacy exposure, and connectivity dependence of cloud LLM serving~\cite{zhou2024survey,liu2024edge}. The natural alternative is to deploy smaller LLMs onto edge System-on-Chip (SoC) platforms that integrate CPU and GPU with
shared unified memory.

Edge SoCs are not miniaturized datacenters. We study three consecutive edge-SoC generations in the NVIDIA Jetson lineage: AGX Xavier~\cite{nvidia_xavier_manual}, AGX Orin~\cite{nvidia_orin_manual}, and AGX Thor~\cite{nvidia_thor_trm_2025}. Together, these platforms span Volta, Ampere, and Blackwell classes of edge GPUs, while preserving the broader Jetson software ecosystem~\cite{nvidia_xavier_manual,nvidia_orin_manual,nvidia_thor_trm_2025}.
Unlike datacenter accelerators, these SoCs share memory between CPU and GPU, operate under tight power and thermal budgets~\cite{sun2020arvr}, and rely on platform-specific dynamic voltage and frequency scaling (DVFS)~\cite{taherin2018tsusc,taherin2015stretch} and telemetry interfaces~\cite{nvidia_thor_trm_2025,nvidia_orin_manual,nvidia_xavier_manual}. These constraints introduce edge-specific bottlenecks that LLM-serving systems designed for datacenters~\cite{kwon2023vllm,patel2024splitwise,zhong2024distserve,stojkovic2024dynamollm} do not encounter, so do not address. These constraints and bottlenecks also change how quantization affects edge LLM behavior. A format that reduces model footprint can still introduce backend-specific compute, memory-traffic, power, and thermal effects that single-precision characterization studies do not expose~\cite{frantar2022gptq,lin2024awq,xiao2023smoothquant,lee2023owq,chee2023quip,tseng2024quip,yao2022zeroquant,tan2024mobilequant,shen2024edgeqat,guo2023olive}.

A growing body of work characterizes LLM inference on edge devices, including single-platform deep dives~\cite{dhar2024empirical,nezami2024generative,ardakani2025llmpi,husom2025sustainable}, mobile-platform benchmarks~\cite{li2024palmbench,laskaridis2024melt,xu2024edgellm}, analytical performance models~\cite{pinnock2025edgeprofiler}, cost--latency--privacy tradeoff analyses~\cite{jang2025edge}, and on-device memory-tier studies~\cite{alizadeh2024llmflash,liu2024mobilellm,yuan2024mobilefm}. Generic edge-AI benchmarks~\cite{baller2021deepedgebench,banbury2021mlperf,mlperfinterface2020,MLPerfPower2025,minott2025benchmarking}
have matured in parallel, as has server-side workload
characterization~\cite{xiang2025servegen,niu2025tokenpowerbench,samsi2023words,chittyvenkata2024llmib,wu2025tokensim,cho2024llmservingsim}. Yet, across this body of prior work, three methodological gaps persist. 
\emph{First,} cross-generational comparisons within the same edge SoC lineage remain limited. Such comparisons are needed to expose how SoC microarchitecture evolution, including GPU generation, CPU
organization, and memory bandwidth, changes LLM inference behavior on edge platforms. 
\emph{Second,} studies typically adopt either a Python-framework path (e.g., HuggingFace Transformers~\cite{wolf2020transformers}) or a C++/GGML path (e.g., \texttt{llama.cpp}~\cite{llama_cpp,ggml}), but rarely both under the same per-prompt, phase-aware timing schema. Without such a common schema, backend differences across tokenization,
prefill, per-token generation, de-tokenization, inter-token latency (ITL), and end-to-end latency remain difficult to compare directly. 
\emph{Third,} timing and hardware telemetry are rarely fused at the same phase boundaries across backends, precisions, and platforms. As a result, studies often report aggregate end-to-end latency, throughput, or run-averaged power, but cannot directly compare; for example, decode-phase power, memory traffic, or energy per token between HuggingFace and \texttt{llama.cpp}. Prefill and decode can stress the SoC differently, but aggregate reporting collapses them into a single per-run number and obscures their distinct contributions, even though the prefill/decode boundary is central to server-side LLM serving studies~\cite{patel2024splitwise,zhong2024distserve,niu2025tokenpowerbench}.

We address these gaps with \emph{Hydra}\footnote{\emph{Hydra} is named after the largest constellation, reflecting the breadth of this characterization: three SoC generations, two backends, 13 models, and five execution formats.}, a common-schema, phase-aware workload characterization framework for LLM inference on edge SoCs. Hydra instruments two structurally different inference backends (HuggingFace Transformers~\cite{wolf2020transformers} and \texttt{llama.cpp}~\cite{llama_cpp} on the GGML tensor
library~\cite{ggml}) to emit per-prompt timing records through a canonical schema, and fuses that timing with high-resolution hardware telemetry from \texttt{tegrastats}~\cite{tegrastats} and the NVIDIA Management Library (NVML)~\cite{nvml}, allowing CPU, GPU, memory, power, energy, and thermal behavior to be attributed to the prefill and decode windows of each prompt. 
Rather than tying the methodology to a single backend, precision, or platform instance, Hydra separates the measurement interface from the execution configuration: new models, precision formats, prompt lengths, and SoC generations can be
added while preserving the same phase-aware analysis pipeline. 
In this paper, we instantiate Hydra across two inference backends, three Jetson generations, and five representative execution formats
(\texttt{bf16},~\texttt{F16},~\texttt{Q8\_0},~\texttt{Q6\_K},~\texttt{Q4\_K\_M}), yielding a unified view that prior single-axis characterizations cannot provide. 

Using Hydra, we conduct a multi-dimensional workload characterization of edge LLM inference. First, we analyze performance through end-to-end latency, decode throughput, phase-level timing, and input/output-length sensitivity. Second, we use phase-aligned telemetry to explain those performance trends through CPU-side runtime behavior, GPU effective utilization, and DRAM traffic. Third, we quantify the deployment cost through power, energy per token, total prompt energy, and thermal behavior.

Our analysis leads to three main findings:

\smallskip
\noindent\textbf{Performance is jointly dependent on the choice of backend, model
    architecture, precision, and sequence length.} End-to-end latency
    captures broad platform trends, but Hydra's phase timing shows that
    backend overhead, quantized decode throughput, and input/output-length
    scaling affect different parts of the inference pipeline.

\smallskip
\noindent\textbf{Resource utilization explains why those performance
    trends occur.} Backend choice changes CPU-side orchestration and GPU
    effective utilization; quantization reduces DRAM traffic without
    necessarily reducing GPU occupancy; and SoC-generation differences make
    raw utilization counters misleading to compare without first applying normalization.

\smallskip
\noindent\textbf{Efficiency does not follow latency or bit-width
    alone.} Lower-bit formats often reduce energy per token, but power and
    thermal behavior depend on the quantization implementation, runtime, and
    platform. Faster configurations may draw more power, yet still spend fewer
    joules per generated token.

This paper makes the following contributions:

\smallskip
\noindent\textbf{Hydra, a cross-backend phase-aware characterization methodology.} Hydra instruments HuggingFace Transformers and
    \texttt{llama.cpp} through a shared per-prompt schema, enabling direct
    comparison of tokenization, prefill, generation, de-tokenization, Time to First Token (TTFT), ITL, and end-to-end timing across structurally different backends.

\noindent \textbf{Phase-aligned full-stack telemetry fusion.} Hydra aligns
    runtime timing with CPU, GPU, memory-controller, power, energy, and
    thermal telemetry, producing canonical per-prompt records that attribute
    system behavior to prefill and decode windows.

\smallskip
\noindent\textbf{A broad cross-generation edge-LLM characterization.} We use
    Hydra to evaluate 13 models across seven families, three SoC
    generations, two inference backends, five representative execution
    formats, and input/output-length sweeps.

\smallskip
\noindent\textbf{A multi-dimensional evaluation of edge LLM deployment behavior.}
    We connect performance, utilization, and efficiency to show how backend
    orchestration, quantization format, SoC-generation behavior, and sequence
    length jointly determine latency, throughput, energy, and thermal costs.
    
\smallskip
\noindent\textbf{An open phase-aware edge-LLM trace corpus.} We release
    Hydra's unified per-prompt traces and analysis artifacts, including
    timing, system-resource, and efficiency aggregates, to support
    reproducible comparison against our cross-platform measurements
    (see the Artifact Appendix for details). 

\smallskip
Hydra's source code and the collected per-prompt trace corpus are available open-source at: 
\begin{itemize}
    \item Public repository:
\url{https://github.com/amirtaherin/hydra}.
    \item Archival copy: Zenodo, DOI \href{https://doi.org/10.5281/zenodo.21844843}{10.5281/zenodo.21844843}.
\end{itemize}

\section{Background and Motivation}
\label{sec:background}

\noindent\textbf{Background.}
Modern generative LLMs are decoder-only Transformers~\cite{vaswani2017attention}
served through a prefill phase, which processes the input prompt and
populates the KV cache, followed by a decode phase, which generates tokens
autoregressively. These phases can stress edge SoCs differently because
edge platforms couple CPU cores, GPU cores, memory controllers, and
power-management logic within a shared memory and thermal envelope
~\cite{nvidia_thor_trm_2025,nvidia_orin_manual,nvidia_xavier_manual}.
Deployment behavior also depends on the inference backend and execution
format. High-level Python frameworks such as HuggingFace
Transformers~\cite{wolf2020transformers} provide broad model coverage,
while lightweight backends such as \texttt{llama.cpp}~\cite{llama_cpp} and
GGML~\cite{ggml} reduce software overhead and enable efficient low-bit execution. We use \emph{execution format} to refer to the
backend-specific numeric representation used for inference, including
16-bit floating-point formats and lower-bit weight-only quantized formats.
Lower-bit formats can reduce model footprint and traffic~\cite{frantar2022gptq,tan2024mobilequant}, but the actual benefits depend
on the backend implementation, dequantization overhead, and memory-system behavior. Finally, edge telemetry itself is platform-specific:
\texttt{tegrastats}~\cite{tegrastats} and NVML~\cite{nvml} expose CPU, GPU,
memory-controller, power, and thermal signals with different availability
and semantics across SoC generations.

\begin{figure}[t]
\centering
\vspace{-0.3cm}
\includegraphics[width=\columnwidth]{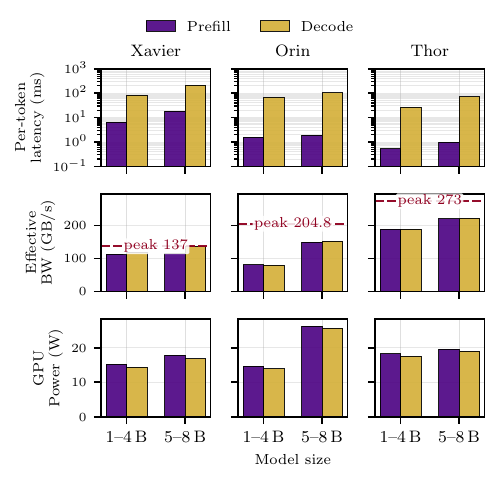}
\vspace{-0.6cm}
\hrule
\caption{\textit{Phase-aware HuggingFace \texttt{bf16} inference across three
consecutive SoC generations. Models are grouped into
1--4\,B and 5--8\,B bins. Rows show per-token latency, effective memory
bandwidth, and GPU power for prefill and decode; dashed lines mark peak
memory bandwidth.}}
\label{fig:motivation}
\vspace{-0.7cm}
\end{figure}

\smallskip
\noindent\textbf{Motivation.}
Recent on-device LLM systems span phones, embedded SoCs, memory-tiered execution, and edge benchmarks~\cite{alizadeh2024llmflash,liu2024mobilellm,
yuan2024mobilefm,liu2024edge}, but aggregate reporting still hides how
performance, resource utilization, and efficiency interact across inference
phases. Fig.~\ref{fig:motivation} illustrates this problem. Along the
\emph{performance} dimension, prefill and decode show different per-token
latency behavior across model sizes and SoC generations. Along the
\emph{resource-utilization} dimension, the same phase maps differently onto
each memory subsystem relative to platform peak bandwidth. Along the
\emph{efficiency} dimension, GPU power does not track latency or bandwidth
uniformly, so the fastest configuration is not necessarily the most
energy-efficient one. These dimensions are coupled: edge LLM behavior depends on SoC generation, backend, execution format, and prompt/output length, but standard
measurements usually expose only one dimension at a time, such as mean latency,
throughput, utilization, or power. Hydra addresses this measurement problem
by treating observability as a stack: token-level timing, phase-level
prefill/decode timing, SoC-level resource signals, and efficiency-level
power, energy, and thermal behavior. Hydra aligns these signals through a
common phase-aware per-prompt schema, enabling the same analysis pipeline
across HuggingFace Transformers and \texttt{llama.cpp}. The rest of the
paper develops this methodology (\S\ref{sec:methodology} \& \S\ref{sec:setup}) and applies it to
evaluate performance, utilization, and efficiency 
(\S\ref{sec:performance}--\S\ref{sec:efficiency}).

\section{Hydra: Phase-Aware Workload Characterization}
\label{sec:methodology}

\begin{figure*}[t!]
\vspace{-0.1cm}
\centering
\includegraphics[width=\textwidth]{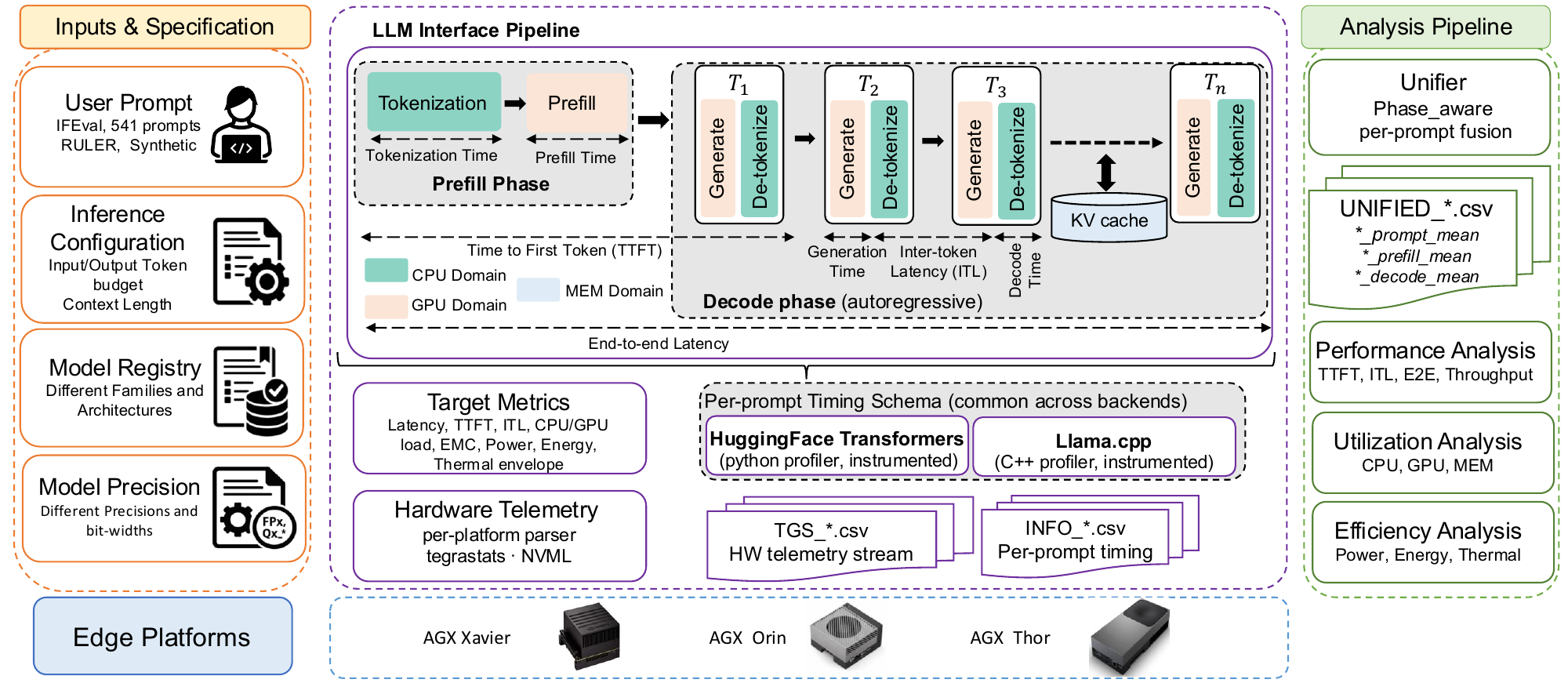}
\vspace{-0.3cm}
\hrule
\vspace{0.1cm}
\caption{\textit{Hydra phase-aware workload characterization workflow. Hydra instruments
HuggingFace Transformers and \texttt{llama.cpp} with a common per-prompt
timing schema, collects SoC telemetry in parallel, aligns both streams per
prompt, and normalizes the fused records into a canonical analysis format.}}
\label{fig:hydra}
\vspace{-0.6cm}
\end{figure*}

Hydra is the measurement methodology for edge-LLM workload characterization. As shown in Fig.~\ref{fig:hydra}, Hydra converts heterogeneous backend execution and platform-specific telemetry into a
common per-prompt record that can be analyzed by phase, backend, precision, and SoC generation. 
This section describes the three components of the
methodology: \textit{(i)} the common timing schema used to instrument
HuggingFace Transformers and \texttt{llama.cpp}; \textit{(ii)} the telemetry
fusion process that aligns runtime events with SoC counters; and
\textit{(iii)} the canonical schema used to normalize platform-specific
signals for downstream performance, utilization, and efficiency analysis.
Section~\ref{sec:setup} then describes how Hydra is instantiated for this
study through the selected SoC generations, model families, workloads, execution formats, and inference configurations.

\smallskip
\noindent\textbf{Measurement Model.}
\label{sec:pipeline}
Hydra uses the standard prefill/decode structure of LLM serving
systems~\cite{patel2024splitwise, zhong2024distserve,
niu2025tokenpowerbench} as a measurement coordinate system. We distinguish
between timing stages and telemetry windows. At the timing level, Hydra
records tokenization, the prefill stage, per-token generation, de-tokenization, ITL, TTFT, and end-to-end latency. At the
telemetry level, Hydra groups these events into two phase windows: a
\emph{prefill phase}, which covers tokenization and the prefill stage, and a
\emph{decode phase}, which covers autoregressive token generation and
de-tokenization until the prompt completes. 
As shown in Fig.~\ref{fig:hydra}, for a prompt with start time $t_\text{start}$, tokenization time $\tau_\text{tok}$, prefill-stage time $\tau_\text{prefill}$, and end time $t_\text{end}$, Hydra defines the prefill-phase window as $[t_\text{start},\, t_\text{start}+\tau_\text{tok}+\tau_\text{prefill}]$, and the decode-phase window as $(t_\text{start}+\tau_\text{tok}+\tau_\text{prefill},\, t_\text{end}]$.
All sampled hardware signals are then aggregated over the prompt, prefill,
and decode windows using the same boundaries for both backends. These shared phase windows are the join point between backend timing records and SoC telemetry, enabling
phase-attributed comparison across backends, execution formats, and
edge-SoC generations.

\smallskip
\noindent\textbf{Instrumentation and Telemetry Fusion.}
\label{sec:architecture}
\label{sec:fusion}
As shown in Fig.~\ref{fig:hydra}, Hydra instruments two structurally
different inference backends: HuggingFace Transformers~\cite{wolf2020transformers}
in Python and \texttt{llama.cpp}~\cite{llama_cpp} in C++ on the GGML tensor
library~\cite{ggml}. Both backends emit one per-prompt timing record using a
common schema for tokenization, prefill, per-token generation, de-tokenization, ITL, TTFT, end-to-end latency, and
start/end timestamps. To avoid measuring asynchronous kernel launch time
instead of completed GPU work, Hydra synchronizes CUDA-dependent regions with
\texttt{torch.cuda.synchronize()} in the Python backend and
\texttt{llama\_synchronize()} in the C++ backend. The \texttt{llama.cpp}
backend also records its internal performance counters for validation.

In parallel, Hydra samples SoC telemetry through
\texttt{tegrastats}~\cite{tegrastats} and NVML~\cite{nvml}. The telemetry
stream captures the platform-exposed CPU, GPU, memory-controller, RAM, power,
and thermal signals. A platform-specific parser converts these raw samples
into timestamped records on the same time base as the runtime profiler. Hydra then fuses backend timing with telemetry on a per-prompt basis. For
each prompt, telemetry samples are aggregated over the full prompt window and
over the prefill and decode phase windows as defined above. The resulting canonical record attaches prompt-level,
prefill-level, and decode-level resource and efficiency statistics to each
timing record. Because \texttt{tegrastats} samples at a fixed wall-clock interval, short
prefill windows often contain only a few samples, while longer decode
windows are sampled much more densely. Hydra emits per-prompt prefill
aggregates as measured; the wider prefill variance in \S\ref{sec:evaluation}
reflects this sampling asymmetry.

\smallskip
\noindent\textbf{Canonical Cross-Platform Schema.}
\label{sec:schema}
Hydra normalizes platform-specific telemetry into a canonical schema so that
the same analysis code can operate across Jetson generations. Raw telemetry
differs by platform: CPU core counts, GPU organization, thermal-zone names,
power-rail labels, and software-stack support are not identical across Xavier,
Orin, and Thor devices. 
Hydra maps these disparate values into shared fields for CPU load,
GPU activity, memory-controller behavior, RAM usage, power, and temperature,
while preserving the source. This normalization is intentionally conservative. Hydra does not hide measurement asymmetries (i.e., some GPU-utilization and power channels
come from different platform interfaces or represent different power domains).
Cross-platform comparisons use the canonical schema for consistent
data access, but the evaluation interprets platform-specific telemetry
differences explicitly when they affect a result.

\smallskip
\noindent\textbf{Validation and Artifact.}
\label{sec:validation}
Because HuggingFace Transformers and \texttt{llama.cpp} use different
execution stacks, Hydra validates the timing definitions rather than assuming
runtime equivalence. The common schema is enforced by construction: shared
columns use identical formulas across both profilers. For \texttt{llama.cpp},
Hydra additionally logs the runtime's built-in performance counters alongside
manual timings for every prompt. Prefill times agree within $0.1$\,ms, and
per-token generation times differ by only $0.06$--$0.08$\,ms. This residual
gap matches the cost of the logit read and greedy \texttt{argmax} included
by Hydra, but excluded from the internal counter, making the difference
bounded and explainable. 
Hydra is released as an open artifact with the profilers, telemetry parsers, unifier, canonical schema, analysis pipeline, and per-prompt record
corpus; the Artifact Appendix details the repository and archival DOI.

\section{Hydra: Workload Characterization Design}
\label{sec:setup}

Section~\ref{sec:methodology} described how Hydra converts backend timing
and SoC telemetry into a common phase-aware record. This section describes
how we instantiate that workflow, shown in Fig.~\ref{fig:hydra}, to
characterize edge LLM inference across the design space. Our experimental
matrix is chosen to disambiguate four effects that are often conflated in edge
LLM studies: SoC generation, model architecture, execution backend/format,
and prompt/output length. We carry out our evaluation on three edge-SoC generations,
a set of instruction-tuned model families, multiple execution
formats across HuggingFace Transformers and \texttt{llama.cpp}, and both
fixed-workload and length-sensitivity prompt sets.

\begin{table}[t]
\centering
\caption{Evaluated edge SoC platforms.}
\label{tab:hardware}
\scriptsize
\setlength{\tabcolsep}{2pt}
\renewcommand{\arraystretch}{1.08}
\begin{tabular}{@{}lccc@{}}
\toprule
\textbf{Feature} & \textbf{Xavier} & \textbf{Orin} & \textbf{Thor} \\
\midrule
GPU arch. & Volta & Ampere & Blackwell \\
GPU SMs/CUDA cores & 8 / 512 & 16 / 2048 & 20 / 2560 \\
CPU & 8-core Carmel & 12-core A78AE & 14-core Neoverse V3AE \\
Memory & 32 GB LPDDR4X & 32 GB LPDDR5 & 128 GB LPDDR5X \\
Peak BW & 137 GB/s & 204.8 GB/s & 273 GB/s \\
L2/system cache & 512 KB / -- & 4 MB / 4 MB & 32 MB / 16 MB \\
Power budget & 30 W & 60 W & 130 W \\
System Software & JP 5.1  & JP 6.2  & JP 7.1 \\
CUDA ver. &  11.4 & 12.6 & 13.2 \\
PyTorch/Transformers & 2.2.0 / 4.46.3 & 2.3.0 / 4.51.3 & 2.11.0 / 5.5.4 \\
\bottomrule
\end{tabular}
\vspace{-0.5cm}
\end{table}
We evaluate three NVIDIA Jetson AGX edge-SoC generations:
Xavier (Volta)~\cite{nvidia_xavier_manual}, Orin (Ampere)~\cite{nvidia_orin_manual}, and Thor (Blackwell)~\cite{nvidia_thor_trm_2025}. These platforms define
one axis of our workload characterization. 
Table~\ref{tab:hardware}
summarizes the compute resources, memory subsystem, power budget,
and software stack of each platform. All three platforms use a unified-memory design: CPU and GPU activity share the same DRAM pool, eliminating explicit
host-to-device transfers in the inference path. Across the three generations, memory capacity
scales by roughly $4\times$ (32~GB $\rightarrow$ 128~GB), peak memory bandwidth by $\sim2\times$ (137 $\rightarrow$ 273~GB/s), while the max power raises more than $4\times$ (30 $\rightarrow$ 130~W). Across these generations, Hydra studies how the evolution of edge-SoC microarchitecture can influence LLM inference behavior.

\subsection{Models and Execution Formats}
\label{sec:setup-models}

The model axis of our characterization covers 13 instruction-tuned,
decoder-only LLMs from seven families: LLaMA, Qwen, Granite, Gemma, Phi,
Mistral, and Moxin. The selected models span 1.24--8.54~B parameters
(Table~\ref{tab:models}), covering both compact sub-2~B models and the
7--8~B range that represents the practical upper end for 32~GB edge SoCs
under 16-bit execution. This range lets Hydra separate effects caused by
model scale from those caused by architecture, such as depth, hidden size,
FFN expansion, attention layout, vocabulary size, and context length. The
\emph{Code} column in Table~\ref{tab:models} defines the short identifiers
used in the x-axis labels of figures throughout the paper. \emph{Mem} is the \texttt{bf16} weight footprint. Q/KV denotes the query and key-value head counts; unequal values indicate grouped-query attention (GQA). Moxin-7B and Gemma-7B exceed the nominal 7B class, but we keep the developer-published names for consistency with public model registries.

The execution-format axis captures how backend and precision choices change
system behavior. We evaluate HuggingFace (HF) Transformers at \texttt{bf16} and
\texttt{llama.cpp} at \texttt{F16}, \texttt{Q8\_0}, \texttt{Q6\_K}, and
\texttt{Q4\_K\_M}. The GGML quantized formats are weight-only
post-training quantization formats at roughly $8$, $6.6$, and $4.5$ bits
per weight, respectively. Because our quantization study is conducted in
\texttt{llama.cpp}, we use \texttt{F16} as the 16-bit quality reference and
report the quality cost of the three GGUF quantized formats relative to
that baseline. Note that Xavier's Volta GPU lacks the native BF16
tensor-core path of Ampere and Blackwell, so Xavier \texttt{bf16} results
reflect backend/software behavior; our
native 16-bit cross-generation comparisons therefore rely on
\texttt{llama.cpp} \texttt{F16}.

Table~\ref{tab:accuracy} summarizes quality changes on WikiText-2
perplexity (PPL) and zero-shot Winogrande~\cite{sakaguchi2021winogrande} and
HellaSwag~\cite{zellers2019hellaswag} accuracy, evaluated with
\texttt{lm-eval-harness}~\cite{eval-harness}. The quality cost is small:
\texttt{Q8\_0} and \texttt{Q6\_K} remain close to \texttt{F16}, while
\texttt{Q4\_K\_M} stays within $1$~PPL point and about $2.4$ accuracy
points in the worst case. Thus, the system-level differences analyzed in
\S\ref{sec:evaluation} primarily reflect runtime, precision, and platform
effects rather than large quality regressions.

\begin{table}[t]
\centering
\caption{Evaluated instruction-tuned LLMs.}
\label{tab:models}
\scriptsize
\setlength{\tabcolsep}{3pt}
\renewcommand{\arraystretch}{1.05}
\resizebox{\columnwidth}{!}{%
\begin{tabular}{@{}lllrcl@{}}
\toprule
\textbf{Code} & \textbf{Model} & \textbf{Family} & \textbf{Params} &
\textbf{Mem} &
\textbf{\begin{tabular}[c]{@{}c@{}}Architecture\\
(L, H, FFN, Q/KV, Attn., Act., Vocab, Ctx.)\end{tabular}} \\
\midrule
LL-1B   & LLaMA-3.2-1B    & LLaMA   & 1.24\,B & 2.36\,GB &
16, 2048, 8192, 32/8, GQA, SwiGLU, 128K, 131K \\
QW-1.5B & Qwen2.5-1.5B    & Qwen    & 1.54\,B & 2.95\,GB &
28, 1536, 8960, 12/2, GQA, SwiGLU, 152K, 32K \\
GE-2B   & Gemma-2B        & Gemma   & 2.51\,B & 4.78\,GB &
18, 2048, 16384, 8/1, MQA, GELU, 256K, 8K \\
GR-2B   & Granite-3.3-2B  & Granite & 2.53\,B & 4.83\,GB &
40, 2048, 8192, 32/8, GQA, SwiGLU, 49K, 131K \\
QW-3B   & Qwen2.5-3B      & Qwen    & 3.09\,B & 5.99\,GB &
36, 2048, 11008, 16/2, GQA, SwiGLU, 152K, 32K \\
LL-3B   & LLaMA-3.2-3B    & LLaMA   & 3.21\,B & 6.13\,GB &
28, 3072, 8192, 24/8, GQA, SwiGLU, 128K, 131K \\
PH-4B   & Phi-3.5-mini    & Phi     & 3.82\,B & 7.29\,GB &
32, 3072, 8192, 32/32, MHA, SiLU, 32K, 131K \\
QW-7B   & Qwen2.5-7B      & Qwen    & 7.62\,B & 14.57\,GB &
28, 3584, 18944, 28/4, GQA, SwiGLU, 152K, 32K \\
MI-8B   & Ministral-8B    & Mistral & 8.02\,B & 15.30\,GB &
36, 4096, 12288, 32/8, GQA, SwiGLU, 131K, 32K \\
LL-8B   & LLaMA-3.1-8B    & LLaMA   & 8.03\,B & 15.32\,GB &
32, 4096, 14336, 32/8, GQA, SwiGLU, 128K, 131K \\
MO-7B   & Moxin-7B        & Moxin   & 8.11\,B & 15.48\,GB &
36, 4096, 14336, 32/8, GQA, SwiGLU, 32K, 32K \\
GR-8B   & Granite-3.3-8B  & Granite & 8.17\,B & 15.58\,GB &
40, 4096, 12800, 32/8, GQA, SwiGLU, 49K, 131K \\
GE-7B   & Gemma-7B        & Gemma   & 8.54\,B & 16.28\,GB &
28, 3072, 24576, 16/16, MHA, GELU, 256K, 8K \\
\bottomrule
\end{tabular}%
}
\vspace{-0.15cm}
\end{table}

\begin{table}[t]
\centering
\caption{Average quality change of each \texttt{llama.cpp} quantized
format relative to \texttt{F16}. Positive $\Delta$PPL is worse; negative
$\Delta$accuracy is worse.}
\label{tab:accuracy}
\scriptsize
\setlength{\tabcolsep}{3.5pt}
\renewcommand{\arraystretch}{1.05}
\begin{tabular}{@{}lrrrrrr@{}}
\toprule
\textbf{Format} &
\textbf{\shortstack{Mean\\$\Delta$PPL}} &
\textbf{\shortstack{Max\\$\Delta$PPL}} &
\textbf{\shortstack{Mean\\$\Delta$Wino.}} &
\textbf{\shortstack{Max\\drop}} &
\textbf{\shortstack{Mean\\$\Delta$Hella.}} &
\textbf{\shortstack{Max\\drop}} \\
\midrule
\texttt{Q8\_0}    & $-0.03$ & $0.08$ & $+0.20$ & $0.47$ & $+0.02$ & $0.21$ \\
\texttt{Q6\_K}    & $+0.01$ & $0.23$ & $-0.17$ & $0.94$ & $-0.01$ & $0.30$ \\
\texttt{Q4\_K\_M} & $+0.36$ & $0.91$ & $-0.78$ & $2.37$ & $-0.43$ & $0.79$ \\
\bottomrule
\end{tabular}
\vspace{-0.5cm}
\end{table}

\subsection{Evaluated Prompts and Sequence Lengths}
\label{sec:setup-workload}

The workload axis is designed to separate standard instruction-following
behavior from controlled input- and output-length effects. The main
characterization corpus uses IFEval~\cite{zhou2023ifeval}, which contains
541 prompts spanning 25 verifiable instruction types. Prompt lengths range
from 13 to 345 tokens (mean 47, std 23). For system measurements, we disable
early stopping and use a fixed 500-token decode budget, ensuring that all
models execute a uniform decode workload rather than stopping at
model-dependent end-of-sequence points. 

The main sweep combines three SoC generations, 13 models, and 5 
execution formats: one HuggingFace format and four \texttt{llama.cpp}
formats. This yields
$3 \times 13 \times (1_{\mathrm{HF}} + 4_{\mathrm{llama.cpp}}) = 195$
(platform, model, execution-format) cells. Five Xavier \texttt{F16}
configurations with 7--8~B models fail to load due to the JetPack~5
cuBLAS/NVMAP interaction discussed in \S\ref{sec:performance}; the remaining
190 cells contribute roughly $103{,}000$ per-prompt records.

To test whether the IFEval trends hold beyond short prompts and fixed
500-token responses, we add two targeted sensitivity corpora:
\begin{itemize}[leftmargin=*, itemsep=0pt, topsep=2pt]
  \item \emph{\textbf{S1}: input-length sweep.} 90 prompts from
    RULER~\cite{hsieh2024ruler}, balanced across three needle-in-a-haystack (NIAH)/tracking task
    families at exact input-length tiers of 1k, 3k, and 5k tokens
    (10 prompts $\times$ 3 tasks $\times$ 3 tiers), with
    fixed 500-token decode budget.
    
  \item \emph{\textbf{S2}: output-length sweep.} 30 IFEval prompts filtered to a
    40--60 token input band and swept across output budgets of 1k, 3k,
    and 5k tokens, isolating decode-length scaling while holding prefill
    cost nearly fixed.
\end{itemize}

The sensitivity corpora run on Orin and Thor using \texttt{llama.cpp}
\texttt{Q4\_K\_M} and HuggingFace \texttt{bf16} for six LLaMA/Qwen models
(LL-1B/3B/8B and QW-1.5B/3B/7B). Xavier is excluded from these long-context
sweeps because 5k-token prompts with 7--8~B models exceed its practical
memory envelope. Together, the sensitivity corpora add roughly $4{,}300$
records, bringing the full study to about $107{,}000$ per-prompt records.

\smallskip
\noindent\textbf{Execution Policy.}
The execution policy is chosen to match single-prompt interactive edge
deployment. All runs use batch size 1, greedy decoding, token-by-token
generation, and KV-cache reuse. Both HuggingFace Transformers and
\texttt{llama.cpp} execute on the integrated GPU through CUDA. For
\texttt{llama.cpp}, we request full GPU offload using
\texttt{n\_gpu\_layers=99}; configurations that cannot allocate under this
policy are marked as failed rather than partially offloaded, preserving a
consistent backend comparison.
\section{Evaluation}
\label{sec:evaluation}

Hydra enables us to analyze edge LLM inference along three connected dimensions:
performance, system-resource utilization, and efficiency. These dimensions map
directly to Hydra's design: the common per-prompt timing schema exposes
latency and throughput behavior; the phase-aligned telemetry fusion explains
how CPU, GPU, and memory activity produce those trends, and the power,
energy, and thermal metrics quantify the deployment cost of each backend,
precision, model, and SoC-generation choice.  
We organize the analysis around three questions. \textbf{Q1}: How do platform generation, backend, model architecture, and precision affect latency? \textbf{Q2}: What CPU, GPU, and memory behaviors explain those trends? \textbf{Q3}: What are the associated power, energy, and thermal costs? This structure connects Hydra's phase-aware measurements to deployment decisions rather than reporting aggregate performance alone.

\subsection{Performance Analysis}
\label{sec:performance}

\begin{figure*}[t]
\centering
\includegraphics[width=\textwidth]{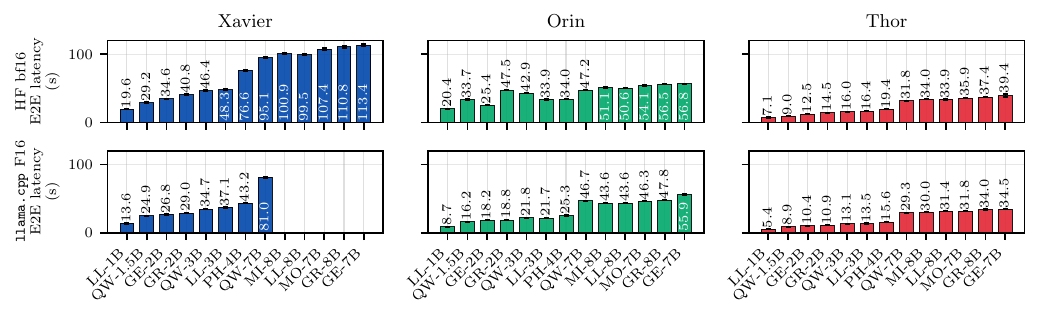}
\vspace{-0.6cm}
\hrule
\vspace{-0.0cm}
\caption{\textit{End-to-end latency across SoC generations and backends. Missing bars indicate failed full-GPU-offload configurations.}}
\label{fig:perf-e2e}
\vspace{-0.6cm}
\end{figure*}

\begin{figure}[t]
\vspace{-0.2cm}
\centering
\includegraphics[width=\columnwidth]{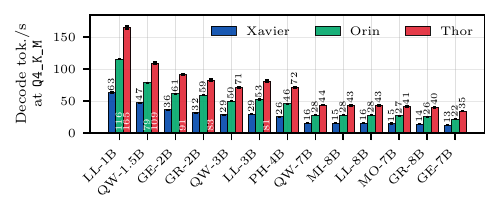}
\vspace{-0.6cm}
\hrule
\vspace{-0.0cm}
\caption{\textit{\texttt{llama.cpp} decode throughput at
\texttt{Q4\_K\_M} across three SoC generations. Quantization improves
decode throughput and compresses the cross-platform gap relative to
16-bit execution.}}
\label{fig:perf-q4}
\vspace{-0.8cm}
\end{figure}

\begin{figure}[t]
\centering
\includegraphics[width=\columnwidth]{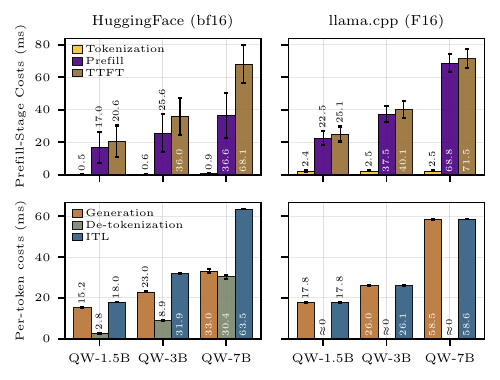}
\vspace{-0.6cm}
\hrule
\vspace{0.1cm}
\caption{\textit{Phase-level latency decomposition on Thor for the Qwen2.5
family under HuggingFace \texttt{bf16} and \texttt{llama.cpp} \texttt{F16}.
Hydra exposes prefill-stage timing, per-token generation, de-tokenization, TTFT, and ITL under the same schema.}}
\label{fig:perf-phases}
\vspace{-0.8cm}
\end{figure}

\begin{figure*}[ht!]
\centering
\includegraphics[width=\textwidth]{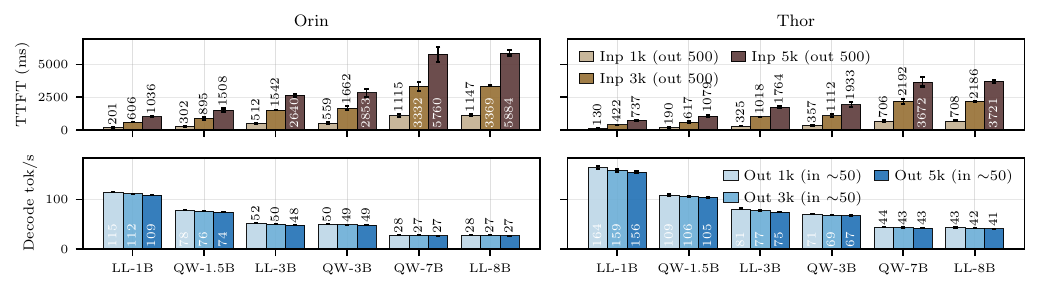}
\vspace{-0.6cm}
\hrule
\vspace{-0.0cm}
\caption{\textit{Input- and output-length sensitivity on Orin and Thor using
\texttt{llama.cpp} \texttt{Q4\_K\_M}. Top: TTFT for 1k/3k/5k-token inputs
with 500 output tokens. Bottom: decode throughput for 1k/3k/5k output
budgets with short inputs.}}
\label{fig:perf-sensitivity}
\vspace{-0.6cm}
\end{figure*}

To answer \textbf{Q1}, we use Hydra's common timing schema to analyze
performance from four views: end-to-end latency across SoC generations and
backends (Fig.~\ref{fig:perf-e2e}), quantized decode throughput (Fig.~\ref{fig:perf-q4}), phase-level latency attribution (Fig.~\ref{fig:perf-phases}), and input/output-length sensitivity (Fig.~\ref{fig:perf-sensitivity}). Together, these views separate overall performance trends from the backend, execution format, model-architecture, and
sequence-length effects that produce them.

\smallskip

\noindent\textbf{End-to-end latency reflects platform, backend, and model-family effects.}
Fig.~\ref{fig:perf-e2e} shows the overall latency trends at similar
16-bit precision. Newer SoC generations reduce latency substantially, but
the gain is not uniform: running HuggingFace \texttt{bf16}, Thor is about
$2.8$--$2.9\times$ faster than Xavier, while the Thor/Orin gap shrinks from
roughly $2.9\times$ on LL-1B to about $1.4\times$ on GE-7B. 
Our backend choice also impacts performance: at similar 16-bit execution, \texttt{llama.cpp} consistently improves end-to-end latency over HuggingFace, with the largest gains on smaller models where runtime overhead is a larger fraction of decode time. Finally, model family matters within the same, similarly sized models: 7--8\,B models show noticeably different latencies, and several
Xavier \texttt{F16} \texttt{llama.cpp} runs fail to allocate memory despite 32\,GB
of unified memory. Thus, end-to-end latency already shows that edge LLM
performance is not determined by parameter count or hardware generation
alone; it depends on the interaction between platform, backend, and model
architecture.

\smallskip
\noindent\textbf{Quantization changes the decode-throughput behavior. \iffalse regime\fi}
Fig.~\ref{fig:perf-q4} shows that \texttt{Q4\_K\_M} substantially increases
decode throughput across all three generations by reducing the weight
traffic per generated token. The effect is large enough to make smaller
models practical (even on Xavier): sub-3\,B models reach roughly
$25$--$60$~tok/s, despite Xavier's older Volta architecture and lower memory
bandwidth. For 7--8\,B models, however, throughput compresses into a much
narrower range: Thor reaches $35$--$44$~tok/s, Orin $22$--$28$~tok/s, and
Xavier $13$--$16$~tok/s. Thus, quantization shifts the deployment boundary,
but it does not eliminate the effects of model scale or SoC generation.

\smallskip
\noindent\textbf{Phase attribution explains backend latency differences.}
The previous figures show that our choice of backend impacts end-to-end latency;
Fig.~\ref{fig:perf-phases} shows why. Hydra reports the same timing stages
for HuggingFace and \texttt{llama.cpp}, separating model execution from
runtime overheads. Running HuggingFace, ITL takes
substantially more time than raw generation because each token step also
pays CPU-side orchestration and de-tokenization costs. For QW-7B on
Thor, generation takes $33.0$\,ms, but ITL reaches $63.5$\,ms; the
additional cost is largely exposed as de-tokenization and per-step runtime
overhead. Under \texttt{llama.cpp}, de-tokenization is sub-millisecond and
ITL nearly matches generation time ($58.6$\,ms vs.\ $58.5$\,ms), indicating
that most per-token latency is spent in the measured generation stage.

This attribution reveals a non-obvious backend tradeoff: \texttt{llama.cpp}
can achieve lower end-to-end latency, even when its raw per-token generation latency is higher than HuggingFace. The advantage comes from lower
runtime overhead, not simply faster model kernels. The tighter TTFT and
prefill variation under \texttt{llama.cpp} further suggest more stable
prompt-side execution, which we revisit in the system-utilization analysis (\S\ref{sec:system}).
Thus, Hydra turns an aggregate backend comparison into a stage-level
explanation of where latency is introduced and where optimization effort
should be directed.

\smallskip
\noindent\textbf{Input and output length stress different performance components.}
Fig.~\ref{fig:perf-sensitivity} separates input-length scaling from
output-length scaling. In the input sweep, TTFT increases almost
proportionally with prompt length: increasing the input tokens from 1k to 5k increases
the TTFT by about $5.0$--$5.7\times$ across the tested models and platforms.
For larger models, this pushes first-token latency into the multi-second
range, making prefill responsiveness the limiting factor for long-context
interactive use. The output sweep shows the opposite behavior. With short inputs fixed, decode throughput remains nearly flat as the output budget increases from 1k to
5k tokens, changing by only a few percent in most cases. Thus, within this
range, longer outputs do not substantially degrade steady-state token
generation, while longer inputs directly increase first-token delay. This
separation is exactly why Hydra treats prefill and decode as distinct
performance components rather than collapsing them into a single latency or
throughput number.

\subsection{System Resource Utilization}
\label{sec:system}

To answer \textbf{Q2}, Hydra shows that the performance trends in
\S\ref{sec:performance} arise from three system mechanisms: CPU-side runtime
orchestration, GPU effective utilization, and DRAM traffic. Our choice of backend
changes how much the CPU performs between token steps; quantization changes
the amount of DRAM traffic needed per generated token; and SoC generation changes how
GPU frequency, memory bandwidth, and utilization interact. These effects are
only visible because Hydra aligns timing and telemetry at the same
per-prompt phase boundaries.

\begin{figure*}[t]
\centering
\includegraphics[width=.85\textwidth]{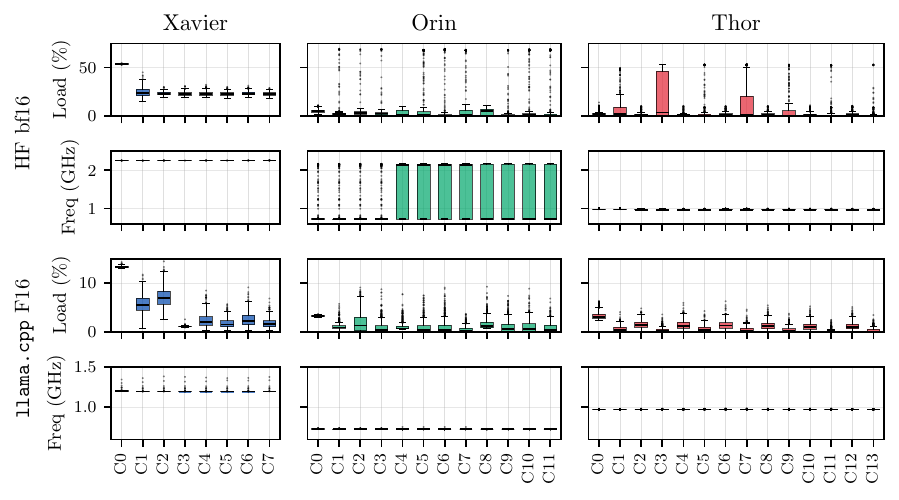}
\vspace{-0.2cm}
\hrule
\vspace{-0.0cm}
\caption{\textit{Per-core CPU load and frequency for Qwen2.5-7B across SoC generations and backends. HuggingFace exposes CPU-side orchestration and DVFS behavior, while \texttt{llama.cpp} largely removes
per-step CPU activity.}}
\label{fig:sys-cpu}
\vspace{-0.5cm}
\end{figure*}

\smallskip
\noindent\textbf{CPU-side behavior explains backend overhead.}
Fig.~\ref{fig:sys-cpu} explains the backend effect observed in
\S\ref{sec:performance}. Running HuggingFace \texttt{bf16}, CPU activity is
visible on every generation: Xavier keeps all cores at peak frequency, Orin
shows scheduler migration across its three clusters, and Thor concentrates
work on a small number of cores, while the rest remain mostly idle. These
patterns show that HuggingFace inference is not only a GPU workload; each
token step also includes CPU-side orchestration, synchronization, and runtime
management overhead.

Running \texttt{llama.cpp} \texttt{F16}, the CPU footprint largely disappears.
The per-core load drops below $15\%$ across all platforms, with most cores near
idle frequency. This is the system-level counterpart of the phase result in
Fig.~\ref{fig:perf-phases}: when the backend removes most per-token CPU
orchestration, ITL aligns closely with generation time and the prompt-stage
variance is reduced. Thus, backend speedups depend on factors beyond
GPU kernels; they also need to account for CPU-side work between token steps.

\begin{table*}[t]
\centering
\caption{Phase-split GPU effective utilization and DRAM effective bandwidth
on Orin for the Qwen2.5 family across backends and execution formats.
Values are mean(std) over prompts.}
\label{tab:gpu-mem-orin}
\setlength{\tabcolsep}{2.5pt}
\renewcommand{\arraystretch}{1.0}
\footnotesize
\begin{tabular}{@{}l l cc cc cc cc cc@{}}
\toprule
& &
\multicolumn{2}{c}{HF bf16} &
\multicolumn{2}{c}{F16} &
\multicolumn{2}{c}{\texttt{Q8\_0}} &
\multicolumn{2}{c}{\texttt{Q6\_K}} &
\multicolumn{2}{c}{\texttt{Q4\_K\_M}} \\
\cmidrule(lr){3-4}
\cmidrule(lr){5-6}
\cmidrule(lr){7-8}
\cmidrule(lr){9-10}
\cmidrule(lr){11-12}
Model & Phase
& $U_\mathrm{eff}$ & $BW_\mathrm{eff}$
& $U_\mathrm{eff}$ & $BW_\mathrm{eff}$
& $U_\mathrm{eff}$ & $BW_\mathrm{eff}$
& $U_\mathrm{eff}$ & $BW_\mathrm{eff}$
& $U_\mathrm{eff}$ & $BW_\mathrm{eff}$ \\
\midrule
\multirow{2}{*}{QW-1.5B} & Pre & 28.7(1.6) & 47.1(1.8) & 86.2(14.6) & 96.0(3.8) & 89.1(12.2) & 85.9(3.2) & 87.4(11.6) & 81.7(2.7) & 88.8(11.6) & 81.8(2.3) \\
 & Dec & 27.3(0.2) & 47.1(0.2) & 97.1(0.1) & 96.2(0.3) & 95.9(0.1) & 86.0(0.4) & 95.3(0.2) & 81.6(0.5) & 94.1(0.3) & 81.7(0.5) \\
\addlinespace
\multirow{2}{*}{QW-3B} & Pre & 42.2(6.6) & 75.5(2.9) & 85.4(11.1) & 140.9(5.8) & 85.4(10.2) & 106.3(4.1) & 84.8(10.8) & 100.1(3.7) & 85.2(12.1) & 100.4(3.5) \\
 & Dec & 38.5(0.4) & 73.9(0.6) & 97.4(0.1) & 141.2(0.3) & 96.8(0.1) & 106.4(0.3) & 96.3(0.1) & 100.3(0.4) & 95.5(0.2) & 100.4(0.5) \\
\addlinespace
\multirow{2}{*}{QW-7B} & Pre & 90.9(5.7) & 151.2(6.3) & 91.1(6.1) & 150.4(6.2) & 91.2(6.7) & 128.6(5.1) & 91.7(6.6) & 123.4(4.9) & 90.1(8.0) & 127.2(5.1) \\
 & Dec & 95.6(0.1) & 152.8(0.3) & 98.2(0.1) & 151.5(0.2) & 97.7(0.1) & 130.2(0.4) & 97.5(0.1) & 124.7(0.3) & 97.1(0.1) & 128.6(0.4) \\
\bottomrule
\end{tabular}
\vspace{-0.35cm}
\end{table*}

\smallskip
\noindent\textbf{GPU and DRAM utilization explain backend and quantization trends.}
Table~\ref{tab:gpu-mem-orin} connects the performance trends from
\S\ref{sec:performance} to GPU and memory behavior on Orin. We report
effective GPU utilization
$U_\mathrm{eff}=\mathrm{load}\times\mathrm{freq}/\mathrm{freq}_\mathrm{peak}$
and effective DRAM bandwidth $BW_\mathrm{eff}$, using Orin's
204.8\,GB/s peak bandwidth. The key comparison is the decode phase, which
dominates end-to-end latency for the 500-token main workload; prefill is included for completeness, but has higher variance because its execution windows are shorter.

The first lesson is that backend choice changes how well the GPU is fed.
For small Qwen models, HuggingFace leaves both GPU utilization and DRAM
bandwidth well below peak, matching the CPU-side orchestration gaps observed
in Fig.~\ref{fig:sys-cpu}. Switching to \texttt{llama.cpp} at the same
16-bit precision, decode utilization increases and is close to saturation. The
effective bandwidth also increases without changing the model. This explains why
\texttt{llama.cpp} improves end-to-end latency in Fig.~\ref{fig:perf-e2e}:
the native runtime reduces CPU-side gaps and keeps the accelerator busier.

The second lesson is that quantization reduces memory pressure without
necessarily reducing GPU occupancy. Across \texttt{llama.cpp} formats,
decode $U_\mathrm{eff}$ remains high, while $BW_\mathrm{eff}$ falls as
weight precision decreases. This explains the throughput behavior in
Fig.~\ref{fig:perf-q4}: quantization improves tokens/s by reducing DRAM
traffic per token, but the GPU can remain highly occupied because the runtime
continues streaming token work efficiently.

The third lesson is that model scale amortizes backend overhead. For QW-7B,
HuggingFace and \texttt{llama.cpp} both result in high GPU utilization and
similar DRAM bandwidth utilization at 16-bit precision. At this size, the amount of per-token model work is large enough that the Python-side overhead becomes 
a smaller portion of the decode step. Thus, the backend performance gap that is large for smaller models narrows as model size increases, matching the end-to-end trend in
Fig.~\ref{fig:perf-e2e}.

\begin{table}[t]
\centering
\caption{HF \texttt{bf16} GPU effective utilization on Orin and Thor
across input- and output-length sweeps for the Qwen2.5 family. Values are
mean(std) over prompts.}
\label{tab:ueff-hf-orin-thor}
\setlength{\tabcolsep}{2pt}
\renewcommand{\arraystretch}{0.95}
\scriptsize
\begin{tabular}{@{}c l l ccc ccc@{}}
\toprule
\multirow{2}{*}{\rotatebox[origin=c]{90}{Model}}
& \multirow{2}{*}{\rotatebox[origin=c]{90}{Platf.}}
& \multirow{2}{*}{\rotatebox[origin=c]{90}{Phase}}
& \multicolumn{3}{c}{Input sweep, Out=500}
& \multicolumn{3}{c}{Output sweep, In=50} \\
\cmidrule(lr){4-6} \cmidrule(lr){7-9}
& & &
1k & 3k & 5k
& 1k & 3k & 5k \\
\midrule

\multirow{4}{*}{\rotatebox[origin=c]{90}{QW-1.5B}}
& Orin & Pre
& 53.7(10.4) & 74.9(4.2) & 80.8(3.7)
& 27.8(4.7) & 28.3(4.9) & 26.2(5.0) \\

& Orin & Dec
& 28.8(0.1) & 30.6(0.3) & 32.8(0.5)
& 27.9(0.1) & 28.6(0.1) & 27.8(0.2) \\

& Thor & Pre
& 63.2(20.8) & 55.5(17.2) & 59.8(27.8)
& 68.2(25.2) & 73.6(21.6) & 70.9(16.2) \\

& Thor & Dec
& 85.4(0.5) & 85.7(0.3) & 86.3(0.3)
& 85.5(0.3) & 85.7(0.1) & 85.9(0.1) \\

\midrule

\multirow{4}{*}{\rotatebox[origin=c]{90}{QW-3B}}
& Orin & Pre
& 79.9(10.0) & 92.0(1.8) & 94.1(1.1)
& 39.8(6.7) & 39.7(7.2) & 38.2(7.0) \\

& Orin & Dec
& 42.7(0.6) & 46.9(2.1) & 52.0(3.0)
& 40.2(0.1) & 41.2(0.2) & 40.8(0.2) \\

& Thor & Pre
& 78.5(18.9) & 87.7(7.9) & 86.7(6.2)
& 74.3(22.1) & 79.4(19.5) & 73.1(20.3) \\

& Thor & Dec
& 88.9(0.3) & 88.9(0.2) & 89.2(0.2)
& 88.9(0.2) & 88.8(0.1) & 88.9(0.1) \\

\midrule

\multirow{4}{*}{\rotatebox[origin=c]{90}{QW-7B}}
& Orin & Pre
& 95.5(11.2) & 97.7(0.7) & 97.9(0.4)
& 90.8(17.3) & 86.0(17.0) & 79.9(21.7) \\

& Orin & Dec
& 95.4(0.2) & 95.6(0.2) & 95.8(0.1)
& 95.4(0.1) & 95.3(0.1) & 95.4(0.1) \\

& Thor & Pre
& 74.2(26.4) & 72.4(13.7) & 73.3(8.5)
& 77.7(19.3) & 76.8(18.4) & 72.9(20.6) \\

& Thor & Dec
& 92.9(0.3) & 93.0(0.2) & 93.2(0.2)
& 92.8(0.2) & 92.9(0.1) & 93.0(0.1) \\

\bottomrule
\end{tabular}
\vspace{-0.45cm}
\end{table}

\smallskip
\noindent\textbf{Length sweeps explain prefill/decode utilization behavior.}
Table~\ref{tab:ueff-hf-orin-thor} connects the length-sensitivity results
from Fig.~\ref{fig:perf-sensitivity} to GPU utilization. The primary
pattern is that input length changes prefill utilization, while output length
does not substantially change decode utilization. On Orin, increasing the
input from 1k to 5k tokens raises prefill $U_\mathrm{eff}$ for smaller Qwen
models, indicating that longer prompts provide enough work to keep the GPU
busy during the prefill phase. In contrast, increasing the output budget
from 1k to 5k tokens leaves decode $U_\mathrm{eff}$ nearly flat, because each
generated token still executes the same per-token loop.

The Orin/Thor contrast also shows why utilization must be interpreted with
platform behavior in mind. Thor maintains high decode $U_\mathrm{eff}$ across
the sweep, while Orin achieves lower utilization on smaller models and
approaches saturation only for QW-7B. This explains the \textbf{Q1} sensitivity result in Fig.~\ref{fig:perf-sensitivity}:
longer inputs increase first-token cost by expanding prefill work, whereas
longer outputs mostly extend steady-state decode without changing the
per-token utilization behavior. Thus, sequence length affects performance
through different system paths, depending on whether it expands the prompt or
the generated output.

\subsection{Efficiency}
\label{sec:efficiency}

To answer \textbf{Q3}, we use Hydra's power, energy, and thermal metrics to
quantify the cost of the performance and utilization trends identified in
\S\ref{sec:performance} and \S\ref{sec:system}. The key question is not only
which configuration is faster, but which configuration produces more tokens
per joule and whether this power and thermal behavior is sustainable.
We analyze decode-phase power, energy per generated token, thermal
response, and total prompt energy across backends, execution formats, model sizes, and SoC generations.

\begin{table}[t]
\centering
\caption{Decode-phase efficiency on Orin and Thor for the Qwen2.5 family.
Each metric cell reports Orin/Thor. Values are mean (std) over prompts;
temperatures in $^\circ$C}.
\label{tab:efficiency-orin-thor}
\setlength{\tabcolsep}{1.1pt}
\renewcommand{\arraystretch}{0.92}
\scriptsize
\begin{tabular}{@{}c l cccc@{}}
\toprule
Model & Pr. 
& Power (W) 
& mJ/tok 
& $T_\mathrm{GPU}$ 
& $T_\mathrm{CPU}$ \\
\midrule

\multirow{5}{*}{\rotatebox[origin=c]{90}{QW-1.5B}}
& bf16 
& 16.2(0.1)/32.8(0.1)
& 1092(4)/591(1)
& 50.9(0.3)/55.9(0.1)
& 54.2(0.4)/55.8(0.5) \\

& F16
& 35.4(0.1)/47.2(0.3)
& 1142(3)/843(2)
& 63.8(0.4)/65.2(0.4)
& 63.9(0.5)/63.6(0.4) \\

& \texttt{Q8} 
& 34.9(0.1)/37.3(0.2)
& 672(2)/561(2)
& 63.1(0.3)/58.3(1.0)
& 63.2(0.3)/56.7(1.0) \\

& \texttt{Q6}
& 42.8(0.1)/55.7(0.5)
& 676(2)/533(2)
& 68.2(0.5)/69.6(0.5)
& 67.2(0.5)/67.0(0.4) \\

& \texttt{Q4}
& 39.9(0.1)/47.0(0.4)
& 509(1)/431(2)
& 63.8(0.5)/62.0(0.6)
& 62.9(0.5)/60.7(0.5) \\

\midrule

\multirow{5}{*}{\rotatebox[origin=c]{90}{QW-3B}}
& bf16
& 20.9(0.2)/34.1(0.1)
& 1788(10)/1088(3)
& 54.0(0.5)/56.7(0.4)
& 56.8(0.5)/55.7(0.5) \\

& F16
& 41.5(0.1)/53.6(0.2)
& 1812(3)/1399(4)
& 65.4(0.4)/69.1(0.2)
& 65.1(0.4)/67.0(0.2) \\

& \texttt{Q8}
& 39.0(0.1)/42.7(0.2)
& 1209(2)/981(3)
& 66.1(0.2)/62.1(0.4)
& 66.1(0.3)/60.4(0.4) \\

& \texttt{Q6}
& 48.5(0.1)/50.4(0.2)
& 1227(3)/1012(2)
& 69.7(0.8)/66.5(0.3)
& 68.3(0.9)/64.5(0.3) \\

& \texttt{Q4}
& 45.0(0.1)/53.5(0.4)
& 903(2)/755(2)
& 67.9(0.3)/67.2(0.5)
& 66.8(0.3)/65.0(0.5) \\

\midrule

\multirow{5}{*}{\rotatebox[origin=c]{90}{QW-7B}}
& bf16
& 39.6(0.1)/37.8(0.1)
& 3731(11)/2399(7)
& 64.9(0.5)/59.5(0.3)
& 66.0(0.6)/58.1(0.4) \\

& F16
& 43.5(0.0)/51.2(0.0)
& 4068(4)/3000(5)
& 67.7(0.3)/67.5(0.2)
& 67.5(0.4)/65.5(0.2) \\

& \texttt{Q8}
& 42.5(0.1)/51.5(0.2)
& 2462(7)/1864(7)
& 65.4(1.0)/66.7(1.0)
& 65.2(1.1)/64.7(1.0) \\

& \texttt{Q6}
& 54.8(0.1)/68.4(0.3)
& 2559(4)/1983(7)
& 70.4(0.3)/74.3(0.8)
& 68.3(0.3)/71.8(0.8) \\

& \texttt{Q4}
& 52.3(0.1)/65.8(0.3)
& 1867(4)/1493(4)
& 71.6(0.3)/74.6(0.3)
& 70.0(0.3)/72.5(0.4) \\

\bottomrule
\end{tabular}
\vspace{-0.35cm}
\end{table}

\smallskip
\noindent\textbf{Quantization reduces energy per token, but power is not
monotonic.}
Table~\ref{tab:efficiency-orin-thor} shows that lower-bit formats reduce
energy per generated token across both Orin and Thor. This follows directly
from \textbf{Q2}: quantization reduces DRAM traffic while maintaining high GPU
occupancy, so each token completes faster and burns less energy. The effect
is most noticeable at \texttt{Q4\_K\_M}, which consistently delivers the lowest
mJ/token among the \texttt{llama.cpp} formats.

Power, however, does not follow bit-width monotonically. \texttt{Q6\_K}
often draws more power and runs hotter than \texttt{Q8\_0} and \texttt{Q4\_K\_M} , despite using fewer bits than \texttt{Q8\_0}. This reinforces a key deployment lesson: quantization format is not only a memory-footprint choice. The format also changes unpacking, scaling, and dequantization work, so power and thermal behavior must be measured rather than inferred from bit-width alone.

\smallskip
\noindent\textbf{Thor spends more watts but fewer joules per token.}
Across matched model/format cells, Thor usually draws more instantaneous
power than Orin, but its higher throughput yields lower energy per token.
This connects \textbf{Q1} and \textbf{Q2}: Thor's newer GPU and memory subsystem increase the token
rate enough to offset the higher power draw. The result is that platform
choice changes the energy operating point, not just the latency operating
point.

\smallskip
\noindent\textbf{Thermals identify the active subsystem, but do not bind
single-stream inference.}
The CPU/GPU temperature rows mirror the utilization trends from
\S\ref{sec:system}. Running HuggingFace on Orin, CPU temperatures can exceed
GPU temperatures, consistent with CPU-side runtime orchestration. Under
\texttt{llama.cpp}, the GPU becomes the warmer component, matching the shift
toward sustained GPU execution. Across all measured single-stream runs,
temperatures remain below throttling-relevant levels, so energy/token is
a more useful constraint than peak temperature.

\begin{table}[t]
\centering
\caption{Total energy per prompt across input- and
output-length sweeps. Values are mean(std) over prompts.}
\label{tab:eff-total-J-sweep}
\setlength{\tabcolsep}{2pt}
\renewcommand{\arraystretch}{0.95}
\scriptsize
\begin{tabular}{c c c ccc ccc}
\toprule
\multirow{2}{*}{\rotatebox[origin=c]{90}{Model}}
& \multirow{2}{*}{\rotatebox[origin=c]{90}{Plat.}}
& \multirow{2}{*}{\rotatebox[origin=c]{90}{Format}}
& \multicolumn{3}{c}{Input sweep, Out=500}
& \multicolumn{3}{c}{Output sweep, In=50} \\
\cmidrule(lr){4-6} \cmidrule(lr){7-9}
& & &
1k & 3k & 5k
& 1k & 3k & 5k \\
\midrule

\multirow{4}{*}{\rotatebox[origin=c]{90}{QW-1.5B}}
& Orin & HF bf16
& 530(2) & 562(4) & 597(8)
& 1034(2) & 3126(10) & 5401(21) \\

& Orin & lcpp Q4
& 273(3) & 314(6) & 365(13)
& 516(2) & 1587(2) & 2723(3) \\

& Thor & HF bf16
& 316(2) & 347(5) & 375(7)
& 604(2) & 1893(3) & 3245(4) \\

& Thor & lcpp Q4
& 232(3) & 264(5) & 296(8)
& 438(2) & 1348(5) & 2291(7) \\

\midrule

\multirow{4}{*}{\rotatebox[origin=c]{90}{QW-3B}}
& Orin & HF bf16
& 884(5) & 956(17) & 1045(32)
& 1687(6) & 5119(13) & 8663(27) \\

& Orin & lcpp Q4
& 482(4) & 542(8) & 604(15)
& 913(2) & 2767(2) & 4660(4) \\

& Thor & HF bf16
& 583(4) & 634(7) & 677(9)
& 1111(3) & 3486(3) & 5961(5) \\

& Thor & lcpp Q4
& 412(3) & 472(8) & 534(15)
& 769(3) & 2369(6) & 4042(8) \\

\midrule

\multirow{4}{*}{\rotatebox[origin=c]{90}{QW-7B}}
& Orin & HF bf16
& 1875(12) & 2035(25) & 2167(30)
& 3637(7) & 11038(17) & 18615(23) \\

& Orin & lcpp Q4
& 994(9) & 1135(20) & 1279(33)
& 1883(7) & 5712(8) & 9735(28) \\

& Thor & HF bf16
& 1272(14) & 1347(13) & 1416(13)
& 2464(4) & 7543(10) & 12805(20) \\

& Thor & lcpp Q4
& 794(9) & 914(16) & 1029(26)
& 1506(6) & 4600(18) & 7810(27) \\

\bottomrule
\end{tabular}
\vspace{-0.45cm}
\end{table}

\smallskip
\noindent\textbf{Total prompt energy is dominated by generated length.}
Table~\ref{tab:eff-total-J-sweep} shows that total energy grows primarily
with output length. Increasing generated tokens from 1k to 5k scales energy
almost proportionally ($5.1$ to $5.4\times$) because decode dominates the run time. Increasing input length also raises energy, but more modestly ($11$ to $34\%$), because prefill is paid once while decode is paid once per generated token. This matches the \textbf{Q1} sensitivity result in Fig.~\ref{fig:perf-sensitivity}: long inputs mainly increase first-token delay, while
long outputs accumulate steady-state decode cost.

\smallskip
\noindent\textbf{Backend, precision, and platform choices compound.}
The total-energy sweep shows that no single axis explains deployment cost.
Moving from HuggingFace \texttt{bf16} to \texttt{llama.cpp}
\texttt{Q4\_K\_M} reduces both runtime overhead and weight traffic; moving
from Orin to Thor improves the token rate at higher instantaneous power. Combined,
these choices can more than double the number of prompts served under the
same energy budget. Thus, the practical efficiency decision is joint:
backend, precision, model size, sequence length, and SoC generation must be
chosen together.

\smallskip
\noindent\textbf{Efficiency takeaway.}
\textbf{Q3} completes the chain from \textbf{Q1} and \textbf{Q2}. Performance gains become deployment
gains only when they reduce energy per useful token. Hydra shows that
quantization, native runtime structure, and newer SoC generations can all
improve that metric, but the power and thermal consequences are
format- and platform-dependent. Aggregate latency alone would miss these
tradeoffs.

\section{Discussion, Limitations, and Future Directions}
\label{sec:discussion}

\noindent\textbf{Practical deployment implications.}
Hydra's characterization translates into concrete deployment guidance:
\begin{itemize}[leftmargin=*, itemsep=0pt, topsep=2pt]
  \item Latency-bound small-model deployments should utilize backends with
    low orchestration overhead;
  \item Energy-bound deployments should focus on mJ/token, rather than
    nominal bit-width, since \texttt{Q4\_K\_M} is often best, while
    \texttt{Q6\_K} can regress;
  \item Power/thermal-capped systems do not always benefit, in terms of
    power, from using lower precision; long-input interactive workloads
    should optimize prefill/TTFT, while long-output workloads should
    optimize decode mJ/token; and
  \item Platform selection must consider both native precision support and
    allocator/runtime constraints.
\end{itemize}
We found these choices can more than double the number of prompts served
under a fixed energy budget.

\smallskip
\noindent\textbf{Measurement overhead and validity.}
Hydra's timing overhead is bounded by the GPU synchronization needed for
correct phase attribution ($<0.1$\,ms per phase). The per-token
$0.06$--$0.08$\,ms logit-read/argmax cost is part of decoding, and our
measurements agree with \texttt{llama.cpp} internal counters ($27.04$ vs.\
$27.00$\,ms prefill). Telemetry is collected out-of-process, and each
per-prompt record is ${\sim}0.4$\,KB. The main threats to validity are the
telemetry sampling asymmetry discussed in \S\ref{sec:methodology}, the
batch-size-one scope, and the platform-specific telemetry gaps covered
below.

\smallskip
\noindent\textbf{Limitations.}
Hydra is designed as a portable characterization methodology, but this
study instantiates it on three NVIDIA Jetson SoCs. The measurement schema
is backend- and platform-extensible, yet the specific performance,
utilization, and efficiency trends reported here reflect the Jetson
software stack, telemetry interfaces, CUDA path, and power-management
policies. Second, our quantization study focuses on weight-only GGUF
formats (\texttt{Q8\_0}, \texttt{Q6\_K}, and \texttt{Q4\_K\_M}); activation
quantization, KV-cache quantization, FP8 execution, and fully integer
pipelines may shift the balance between memory traffic, dequantization
work, and energy. Third, we evaluate single-stream, batch-size-one
interactive inference. Continuous batching, concurrent requests,
speculative decoding, and serving-style scheduling can change both
throughput and resource contention. Fourth, our model set is limited to
dense instruction-tuned LLMs in the 1--8\,B range; mixture-of-experts,
multimodal, and larger dense models may expose different memory and
runtime behavior. Finally, power telemetry is limited by the rails exposed
on each platform. Hydra reports the available GPU, CPU/SoC, and memory/IO
rails consistently, but none of the evaluated boards exposes a clean
DRAM-only power rail comparable to a discrete memory power sensor.

\smallskip
\noindent\textbf{Future directions.}
Hydra's phase-aware schema enables several follow-up studies. First,
adaptive runtime control can use per-phase GPU utilization, memory
bandwidth, energy, and thermal headroom to change precision, offload
policy, or scheduling decisions during prefill and decode rather than once
per request; this is not directly supported by current server-oriented LLM
serving systems~\cite{kwon2023vllm,patel2024splitwise,zhong2024distserve}.
Second, KV-cache management is a natural extension: our output-length
sweeps show that total energy grows with generated length, motivating
KV-cache compression, eviction, or precision adaptation for long
generations. Third, Hydra can be extended beyond Jetson-class SoCs by
adding platform-specific telemetry parsers while preserving the canonical
per-prompt schema. This would enable direct comparison across AMD Ryzen AI,
Qualcomm Snapdragon, Apple M-series, and emerging RISC-V edge accelerators.
Finally, extending the current single-stream sweeps to longer contexts and concurrent serving would result in KV-cache fragmentation, memory-capacity pressure, and DRAM-contention, which is future work. 
\section{Related Work}
\label{sec:related}

We position Hydra against prior LLM characterization and benchmarking work
along the dimensions that matter for edge workload characterization:
hardware coverage, backend support, model and precision scope,
phase-aware timing, SoC telemetry, efficiency metrics, and released
artifacts. Table~\ref{tab:related} summarizes the comparison, where: 
\emph{Common schema} means that multiple inference backends are
instrumented using the same per-prompt, phase-aware timing and telemetry
schema, \emph{Phase} denotes whether metrics are attributed separately to
prefill and decode windows.

\smallskip
\noindent\textbf{Edge LLM characterization.}
Recent edge LLM studies span mobile devices, Raspberry Pi-class systems,
Jetson platforms, and analytical models~\cite{dhar2024empirical,
nezami2024generative,ardakani2025llmpi,husom2025sustainable,
jang2025edge,pinnock2025edgeprofiler}. The closest edge-side comparisons
are MELTing Point~\cite{laskaridis2024melt} and
PalmBench~\cite{li2024palmbench}: both evaluate compressed or mobile LLM
execution across edge-class hardware and include runtime, energy, or
thermal measurements. Other studies motivate on-device LLM deployment
through memory-tiering, mobile model design, speculative decoding, and
foundation-model firmware paths~\cite{alizadeh2024llmflash,
liu2024mobilellm,yuan2024mobilefm,xu2024edgellm}. These efforts established
the importance of edge LLM measurement, but did not combine
HuggingFace and \texttt{llama.cpp} under one per-prompt phase-aware schema
with CPU/GPU/memory, power, energy, and thermal telemetry across multiple
edge-SoC generations.

\smallskip
\noindent\textbf{Phase-aware serving and benchmarking.}
Server-side characterization has shown the value of separating prefill and
decode. TokenPowerBench~\cite{niu2025tokenpowerbench} measures phase-level
power on H100s, while Splitwise~\cite{patel2024splitwise},
DistServe~\cite{zhong2024distserve}, \texttt{vLLM}~\cite{kwon2023vllm}, and
ELLIE~\cite{fan2025ellie} motivate phase-aware serving and KV-cache
management. Trace generation,
simulation, and server benchmarks consider request dynamics,
scheduling, and accelerator behavior in datacenters~\cite{xiang2025servegen,
wu2025tokensim,cho2024llmservingsim,samsi2023words,
stojkovic2024dynamollm,kakolyris2025throttllem,chittyvenkata2024llmib}. Hydra offers a richer picture: measured single-device edge SoCs, where shared memory bandwidth,
CPU--GPU coordination, DVFS, and thermal headroom shape execution.

\smallskip
\noindent\textbf{Broader benchmarks, surveys, and quantization.}
Generic edge and ML benchmarking efforts (e.g., the MLPerf suites) standardize top-line latency,
throughput, power, and model-size reporting across devices and workloads
~\cite{mlperfinterface2020,banbury2021mlperf,MLPerfPower2025,
liu2024edge,baller2021deepedgebench,minott2025benchmarking,
zheng2024edgellm-survey}. In parallel, post-training and hardware-aware
quantization methods reduce LLM footprint and improve deployability
~\cite{frantar2022gptq,lin2024awq,lee2023owq,xiao2023smoothquant,
chee2023quip,tseng2024quip,egiazarian2024extreme,yao2022zeroquant,
guo2023olive,ramachandran2025microscopiq,tan2024mobilequant,
shen2024agilequant,shen2024edgeqat}. Hydra is complementary to these
efforts: it does not propose a new quantizer or benchmark score, but
characterizes how deployed execution formats affect timing, resource
utilization, energy, and thermals on edge SoCs.

\smallskip
\noindent\textbf{Hydra's position.}
Hydra connects pieces that prior work typically studies separately:
cross-generation edge SoC behavior, dual-backend instrumentation
(HuggingFace Transformers and \texttt{llama.cpp}~\cite{llama_cpp}), 16-bit
and GGUF quantized execution formats, phase-attributed telemetry, and an
open per-prompt trace corpus. This combination is the key distinction: the
same canonical record links token-level timing to system-resource behavior
and efficiency, allowing backend, precision, model-family, and SoC-generation
effects to be compared within one workload-characterization framework.

\begin{table}[t]
\centering
\caption{Hydra relative to other LLM characterization/benchmarking tools.
Columns use no/part/full to denote whether none, some, or all submetrics
 are reported.}
\label{tab:related}
\scriptsize
\setlength{\tabcolsep}{1.4pt}
\renewcommand{\arraystretch}{1.02}
\resizebox{\columnwidth}{!}{%
\begin{tabular}{l cc ccc ccc cccc cc ccc}
\toprule
&
\multicolumn{2}{c}{\textbf{Hardware}} &
\multicolumn{3}{c}{\textbf{Backend}} &
\multicolumn{3}{c}{\textbf{Models}} &
\multicolumn{4}{c}{\textbf{Coverage}} &
\multicolumn{2}{c}{\textbf{Reporting}} &
\multicolumn{3}{c}{\textbf{Artifact}} \\
\cmidrule(lr){2-3}
\cmidrule(lr){4-6}
\cmidrule(lr){7-9}
\cmidrule(lr){10-13}
\cmidrule(lr){14-15}
\cmidrule(lr){16-18}
\textbf{Work}
 & \rotatebox{90}{Edge}
 & \rotatebox{90}{Cross-gen.}
 & \rotatebox{90}{llama.cpp}
 & \rotatebox{90}{HF}
 & \rotatebox{90}{Comm. sch.}
 & \rotatebox{90}{\# models}
 & \rotatebox{90}{\# families}
 & \rotatebox{90}{Cross-arch}
 & \rotatebox{90}{Precision}
 & \rotatebox{90}{Timing}
 & \rotatebox{90}{HW perf.}
 & \rotatebox{90}{Efficiency}
 & \rotatebox{90}{Phase}
 & \rotatebox{90}{Per-prompt}
 & \rotatebox{90}{Tool}
 & \rotatebox{90}{Dataset}
 & \rotatebox{90}{\# metrics} \\
\midrule

Dhar~\cite{dhar2024empirical}
& yes & no
& yes & no & no
& 1 & 1 & no
& part & part & part & no
& no & no
& no & no & 5 \\

Nezami~\cite{nezami2024generative}
& yes & no
& yes & no & no
& 8 & 7 & yes
& part & part & part & no
& part & no
& yes & no & 4 \\

PalmBench~\cite{li2024palmbench}
& yes & yes
& yes & no & no
& 10 & 8 & yes
& full & part & full & part
& part & no
& yes & no & 7 \\

LLMPi~\cite{ardakani2025llmpi}
& yes & no
& yes & no & no
& 7 & 4 & no
& full & part & no & part
& no & no
& no & no & 6 \\

Husom~\cite{husom2025sustainable}
& yes & no
& yes & no & no
& 4 & 3 & no
& part & part & no & part
& no & part
& yes & no & 4 \\

Jang \& Mor.~\cite{jang2025edge}
& yes & no
& -- & -- & --
& 17 & 4 & no
& -- & part & no & part
& no & no
& no & no & 4 \\

EdgeProf.~\cite{pinnock2025edgeprofiler}
& no & no
& no & no & no
& 4 & 4 & no
& part & part & no & part
& no & no
& yes & no & 4 \\

MELT~\cite{laskaridis2024melt}
& yes & no
& yes & no & no
& 7 & 5 & yes
& part & part & full & full
& full & yes
& yes & no & 7 \\

\midrule

ServeGen~\cite{xiang2025servegen}
& no & no
& -- & -- & no
& 10 & 3 & no
& -- & part & no & no
& no & yes
& yes & yes & 4 \\

TokenPow.~\cite{niu2025tokenpowerbench}
& no & no
& no & yes & no
& 15 & 4 & yes
& part & part & no & part
& full & part
& yes & no & 5 \\

TokenSim~\cite{wu2025tokensim}
& no & no
& no & no & no
& 2 & 2 & no
& part & part & no & no
& part & no
& yes & no & 2 \\

LLMSrvSim~\cite{cho2024llmservingsim}
& no & no
& no & no & no
& 6 & 2 & no
& -- & no & part & no
& part & no
& yes & no & 2 \\

LLM-IB~\cite{chittyvenkata2024llmib}
& no & no
& yes & no & no
& 8 & 3 & yes
& part & part & no & part
& part & no
& yes & yes & 5 \\

\midrule

\textbf{Hydra (ours)}
& \textbf{yes} & \textbf{yes}
& \textbf{yes} & \textbf{yes} & \textbf{yes}
& \textbf{13} & \textbf{7} & \textbf{yes}
& \textbf{full} & \textbf{full} & \textbf{full} & \textbf{full}
& \textbf{full} & \textbf{yes}
& \textbf{yes} & \textbf{yes} & \textbf{16} \\

\bottomrule
\end{tabular}%
}
\vspace{-0.55cm}
\end{table}

\section{Conclusion}
\label{sec:conclusion}

This paper presented \emph{Hydra}, a common-schema, phase-aware workload characterization methodology for LLM inference on edge SoCs. Hydra aligns backend timing from HuggingFace Transformers and \texttt{llama.cpp} with system telemetry, enabling performance, utilization, and efficiency to be attributed to prefill and decode windows. Across three Jetson generations, thirteen LLMs, five execution formats, and input/output-length sweeps, Hydra shows that edge LLM behavior cannot be explained by model size, precision, backend, or platform generation alone. Backend structure changes where latency is introduced, quantization reduces memory traffic and energy, but does not predict power monotonically, and SoC generation changes how utilization and efficiency should be interpreted. These results show that practical edge LLM deployment requires phase-aware, backend-aware, and platform-aware observability rather than aggregate latency/throughput alone. We release Hydra and its per-prompt trace corpus to support reproducible characterization and future edge-LLM research.

\section*{Acknowledgment}
We would like to thank Matin Raayai Ardakani for his assistance during artifact evaluation and anonymous reviewers for their constructive feedback. This work was partially supported by the U.S. National Science Foundation (NSF) SaTC program under Grant No.~2414652, the EU Project dAIEDGE (GA Nr 101120726), and the Innovate UK Horizon Europe Guarantee (GA Nr 10090788). In preparing this paper, the authors used generative AI tools (OpenAI ChatGPT and Anthropic Claude) for language and presentation refinement. All technical content, results, and claims were produced and verified by the authors, who take full responsibility for the content of this paper.

\bibliographystyle{IEEEtran}
\bibliography{reference}

\clearpage
\raggedend
\section*{Artifact Appendix}
\label{sec:artifact}

\subsection*{A.1 Abstract}
This artifact provides \textbf{\textit{two}} research objects: 
\begin{enumerate}
    \item \textbf{Hydra (code)}: the common-schema, cross-backend profilers (HuggingFace Transformers and \texttt{llama.cpp}), the \texttt{tegrastats}/NVML fusion pipeline, the canonical cross-platform schema, and the analysis pipeline.
    \item \textbf{The Hydra corpus (dataset)}: the complete released measurement corpus of $107{,}110$ per-prompt records (286 unified CSVs: 190 main-corpus configurations plus the S1/S2 length-sensitivity sweeps), shipped inside the repository as a split \texttt{xz} tarball.
\end{enumerate}

The artifact is assessed through \textbf{\textit{two}} evaluations with distinct goals:
\begin{enumerate}
    \item \textbf{Evaluation~I} reproduces every computational result of the paper (Figs.~1, 3--7 and Tables~4--7) from the released corpus on any Linux or macOS machine---no GPU or Jetson hardware required.
    \item \textbf{Evaluation~II} validates the measurement pipeline itself: it re-measures the paper's flagship model on the three Jetson testbeds (SSH access provided) and compares fresh measurements against the corpus.
\end{enumerate}

\subsection*{A.2 Artifact Meta-Information Checklist}
\begin{itemize}[leftmargin=*, itemsep=0pt, topsep=2pt]
  \item \textbf{Program:} Hydra profilers (Python + C++), unifier, analysis
    pipeline (Python).
  \item \textbf{Dataset:} $107{,}110$ per-prompt records; 25 timing fields
    + phase-attributed telemetry aggregates (349--419 columns, depending
    on platform/backend).
  \item \textbf{Hardware (Eval.~I):} any Linux or macOS machine (x86-64 or
    ARM); no GPU. Can also run on the provided boards (never concurrently
    with an Evaluation~II measurement), though a separate machine is
    preferred.
  \item \textbf{Hardware (Eval.~II):} NVIDIA Jetson AGX Xavier / Orin /
    Thor; SSH access to all three provided during evaluation.
  \item \textbf{Software (Eval.~I):} Python $\geq$3.10 with
    \texttt{pandas}, \texttt{numpy}, \texttt{matplotlib},
    \texttt{seaborn}.
  \item \textbf{Software (Eval.~II):} pre-provisioned on the boards;
    per-platform Python/PyTorch stacks are documented in
    \texttt{docs/ENVIRONMENT.md}.
  \item \textbf{Metrics:} latency (TTFT, ITL, E2E), throughput, GPU/CPU
    utilization, DRAM bandwidth, power, energy/token, temperature.
  \item \textbf{Output:} paper Figs.~1, 3--7 (PDF) and Tables~4--7
    (\texttt{md}/\texttt{csv}); spot-check comparison report.
  \item \textbf{Disk space:} $\sim$1\,GB (Eval.~I).
  \item \textbf{Time:} Eval.~I $\sim$5\,min; Eval.~II $\sim$0.5--1\,h per
    board (longer on Xavier).
  \item \textbf{Publicly available:} yes (GitHub + archival DOI).
  \item \textbf{Badges applied for:}
  \begin{itemize}
    \item \textbf{Datasets}: Available, Reviewed, Reproducible (all via Eval.~I)
    \item \textbf{Code} Available (Eval.~I), Reviewed
    (Eval.~I: analysis pipeline; Eval.~II: measurement pipeline), Reproducible (Eval.~II).
  \end{itemize} 
\end{itemize}

\subsection*{A.3 Access}
Public artifact repository:
\url{https://github.com/amirtaherin/hydra}. Archival copy:
Zenodo, DOI \href{https://doi.org/10.5281/zenodo.21844843}{10.5281/zenodo.21844843}. The corpus ships \emph{inside} the repository under
\texttt{data/unified/}; the expected outputs ship with the
repository under \texttt{expected\_results/}; per-platform environment recipes and the
archived PyTorch wheels are documented in \texttt{docs/ENVIRONMENT.md}.

\subsection*{A.4 Evaluation I: Reproducing the Paper's Results
(no Jetson GPU needed)}
\textbf{Goal:} regenerate every computational result of the paper from the
released corpus, demonstrating that the \textbf{dataset} is complete and that the
\textbf{analysis pipeline} (corpus loader, canonical cross-platform
schema, and the figure and table generators) runs and reproduces the
published figures and tables. This
evaluation supports the \emph{Available} and \emph{Reviewed} badges for
both research objects and the \emph{Reproducible} badge for the dataset.

\noindent All steps from scratch, on any Linux or macOS machine:
\begin{cmd}
$ git clone https://github.com/amirtaherin/hydra.git
$ cd hydra
$ python3 -m venv .venv && . .venv/bin/activate
$ pip install pandas numpy matplotlib seaborn
$ bash scripts/ae_reproduce.sh
\end{cmd}
The driver first \textbf{verifies the corpus}: a sha256 check of the
release tarball, then an integrity manifest---$107{,}110$ per-prompt
records in 286 unified CSVs (190 main-corpus configurations plus the
S1/S2 sweeps)---that hard-stops on any mismatch. It
then renders Figs.~1 and 3--7 and regenerates Tables~4--7, emitted both as
\texttt{.md} (for visual comparison against the published tables) and
\texttt{.csv} (machine-readable). All regenerated figures and tables are
written to \texttt{ae\_output/} (\texttt{figures/} and \texttt{tables/}
subdirectories).
\vspace{0.5em}

\noindent\textbf{Expected outcome:} \textit{(i)} the manifest prints
\texttt{CORPUS VERIFICATION PASSED} (automated); \textit{(ii)} six figure
PDFs render (observed); \textit{(iii)} the driver's final step reports
that every value in the regenerated tables is identical to the reference
tables shipped in \texttt{expected\_results/} (automated). The same check can be run by hand:
\begin{cmd}
$ diff -r ae_output/tables \
          expected_results/tables
\end{cmd}
\texttt{diff} \emph{prints nothing when the tables are identical}---empty output is the expected success result. The whole evaluation takes
about five minutes on a laptop. \emph{On the provided boards}, skip the clone/venv/pip steps
above---the pre-installed environment activates on login and
\texttt{\char`~/hydra} is already checked out; simply run the driver. Do
NOT run Evaluation~I on a board while an Evaluation~II measurement is in
progress there: the analysis load perturbs the telemetry being recorded.

\subsection*{A.5 Evaluation II: Validating the Measurement Pipeline
(Jetson testbeds, SSH)}
\textbf{Goal:} certify that the \textbf{measurement pipeline}---the
HuggingFace and \texttt{llama.cpp} profilers, the telemetry collection,
and the timing+telemetry unifier---works end-to-end on real hardware,
by re-measuring the paper's
flagship model and comparing fresh measurements against the released
corpus across all three SoC generations. This evaluation supports the
\emph{Reviewed} and \emph{Reproducible} badges for the code.

\smallskip
\noindent\textbf{(a) Spot-check on the provided boards.} Everything is
already set up on the boards: the Python environment activates at login,
the profiler is pre-compiled, and the models are pre-downloaded; the
login banner summarizes these steps:
\begin{cmd}
$ ssh <board>        # credentials via HotCRP
$ cd ~/hydra
$ tmux new -s ae     # keeps the run alive if SSH drops
$ bash scripts/ae_quick.sh
\end{cmd}
(\texttt{tmux} basics: detach with \texttt{Ctrl-b} then \texttt{d};
reattach later with \texttt{tmux attach -t ae}; the run continues while
detached.) The spot-check measures Qwen2.5-7B under HF \texttt{bf16} and
\texttt{llama.cpp} \texttt{Q8\_0}/\texttt{Q6\_K}/\texttt{Q4\_K\_M}
($\sim$20 IFEval prompts, 500-token decode), fuses timing with telemetry,
and prints the fresh decode power, ITL, energy/token, and throughput
next to the corresponding values from the released corpus. 
\vspace{0.5em}

\noindent\textbf{Expected outcome:} the run produces
a fresh mini-corpus (20 prompts $\times$ 4 configurations, unified
exactly like the released corpus) and a comparison table that ends in
\texttt{Result: INVARIANTS PASS}. Success is judged by the two
reproducibility checks described in \S~A.6. A run takes about 30--60
minutes per board; Xavier, the oldest platform, takes somewhat longer.

\smallskip
\noindent\textbf{(b) Rebuilding the profiler from source} (to verify the
build, or on one's own Jetson). This is the only step that requires the
pinned \texttt{llama.cpp} submodule:
\begin{cmd}
$ git submodule update --init --recursive
$ bash scripts/build_llamacpp_profiler.sh
$ bash scripts/ae_quick.sh   # re-run with the rebuilt binary
\end{cmd}
The build script detects the platform and selects the CUDA architecture
(Xavier \texttt{sm\_72}, Orin \texttt{sm\_87}, Thor \texttt{sm\_110}); it
requires \texttt{cmake}, the CUDA toolkit (\texttt{nvcc} on
\texttt{PATH}), and a C++17 compiler. The repository README documents
overrides and troubleshooting; \texttt{docs/ENVIRONMENT.md} gives the
full per-board environment recipes. Full-scale collection
(\texttt{scripts/run\_\{hf,llamacpp\}\_experiments.sh}) takes days of
device time and is not expected of reviewers.

\subsection*{A.6 Interpreting Results}
For \textbf{Evaluation~I}, the driver performs two automated checks. It first
verifies the released corpus itself (the checksum and record manifest of
\S~A.4), and in its final step it verifies that the regenerated tables
are identical to the reference tables shipped in
\texttt{expected\_results/}---this must hold exactly, since the
analysis pipeline is deterministic. Two further comparisons are manual,
made by the reviewer against the paper: the regenerated \texttt{.md}
tables can be compared with the published Tables~4--7---they match at
the printed precision, though in a handful of cells the last printed
digit differs by one (e.g., 86.3 vs.\ 86.2) due to rounding during
manuscript preparation---and the generated figures should look
identical to the published Figs.~1 and 3--7 (same scripts, same data;
reference copies in \texttt{expected\_results/figures/}; compared
visually, not with \texttt{diff}, since PDF files are never
byte-identical across systems).

\smallskip
In \textbf{Evaluation~II}, the comparison applies two reproducibility checks to the
freshly measured mini-corpus. \textit{(1) Quantitative agreement:} every
fresh value is printed next to its corpus reference with the percentage
deviation; values typically fall within $\pm15\%$ of the corpus means
(thermal state and background load shift absolute numbers, so this band is
guidance, not a hard gate). \textit{(2) Qualitative findings}---the
binding pass/fail criterion, printed as \texttt{Qualitative invariants}:
three findings of the paper must reproduce on every board:
\begin{itemize}[leftmargin=*, itemsep=0pt, topsep=2pt]
  \item \texttt{Q6\_K} draws more decode power than \texttt{Q8\_0}
    (bit-width non-monotonicity, \S\ref{sec:efficiency});
  \item \texttt{Q4\_K\_M} has the lowest energy per token among the
    \texttt{llama.cpp} formats (energy efficiency, \S\ref{sec:efficiency});
  \item \texttt{llama.cpp} ITL $<$ HF ITL (runtime-overhead gap,
    \S\ref{sec:performance}).
\end{itemize}
Together, the two evaluations exercise the artifact's full pipeline:
Evaluation~II covers the collection and fusion stages of Fig.~2, and
Evaluation~I the analysis stage. The remaining figure and tables are not covered because they are not
produced by Hydra: Fig.~2 is a drawn diagram, Tables~1, 2, and 8 are
hand-written summaries, and Table~3 comes from the public
\texttt{lm-eval-harness} tool, independent of the telemetry corpus.

\subsection*{A.7 Customization}
The repository follows the paper's architecture: \texttt{inputs/} (prompts,
model registry), \texttt{inference/} (HF profiler, \texttt{llama.cpp}
profiler, telemetry collection), and \texttt{analysis/} (unifier, figure and
table generators). New models, precisions, or prompt sets require only edits
under \texttt{inputs/}; new platforms require a \texttt{tegrastats} parser
(\texttt{inference/telemetry/}). Beyond the paper's figures,
\texttt{analysis/main.py} exposes further plot families (distributions,
scaling, thermal, memory pressure) over the same corpus, e.g.:
\begin{cmd}
$ python3 -m analysis.main --all \
    --results-root <corpus> --out figs/
\end{cmd}

\end{document}